\documentclass[twocolumn,prb,superscriptaddress,aps,showpacs]{revtex4}

\usepackage{graphicx}
\usepackage{dcolumn}
\usepackage{amsmath}
\usepackage{tikz}

\newlength{\figwidth}
\newlength{\figwidthb}
\newcommand{\be}{\begin{equation}}
\newcommand{\ee}{\end{equation}}

\newcommand{\br}{{{\bf{r}}}}

\newcommand{\bea}{\begin{eqnarray}}
\newcommand{\eea}{\end{eqnarray}}
\newcommand{\beal}{\begin{align}}
\newcommand{\eeal}{\end{align}}
\newcommand{\ra}{\rangle}
\newcommand{\la}{\langle}

\newcommand{\dg}{{\dagger}}
\newcommand{\pdg}{{\phantom\dagger}}

\begin{document}

\title{Determination of Hund's coupling in $5d$ Oxides using Resonant Inelastic X-ray Scattering}
\author{Bo Yuan}
\affiliation{Department of Physics, University of Toronto, Toronto, Ontario M5S~1A7, Canada}
\author{J. P. Clancy}
\affiliation{Department of Physics, University of Toronto, Toronto, Ontario M5S~1A7, Canada}
\author{A. M. Cook}
\affiliation{Department of Physics, University of Toronto, Toronto, Ontario M5S~1A7, Canada}
\affiliation{Department of Physics, University of Zurich, Winterthurerstrasse 190, 8057 Zurich, Switzerland}
\author{C. M. Thompson}
\altaffiliation{Department of Chemistry, Purdue University, 560 Oval Drive, West Lafayette, Indiana 47907-2084, USA}
\affiliation{Department of Chemistry and Chemical Biology, McMaster University, Hamilton, Ontario L8S~4L8, Canada}
\author{J. Greedan}
\affiliation{Department of Chemistry and Chemical Biology, McMaster University, Hamilton, Ontario L8S~4L8, Canada}
\affiliation{Brockhouse Institute for Materials Research, McMaster University, Hamilton, Ontario L8S~4L8, Canada}
\author{G. Cao}
\affiliation{Department of Physics, University of Colorado-Boulder, Boulder, Colorado 80309, USA}
\author{B. C. Jeon}
\affiliation{Center for Correlated Electron Systems, Institute for Basic Science (IBS), Seoul 08826, Republic of Korea}
\affiliation{Department of Physics \& Astronomy, Seoul National University, Seoul 08826, Republic of Korea}
\author{T. W. Noh}
\affiliation{Center for Correlated Electron Systems, Institute for Basic Science (IBS), Seoul 08826, Republic of Korea}
\affiliation{Department of Physics \& Astronomy, Seoul National University, Seoul 08826, Republic of Korea}
\author{M. H. Upton}
\affiliation{XOR, Advanced Photon Source,
Argonne National Laboratory, Argonne, Illinois 60439, USA}
\author{D. Casa}
\affiliation{XOR, Advanced Photon Source,
Argonne National Laboratory, Argonne, Illinois 60439, USA}
\author{T. Gog}
\affiliation{XOR, Advanced Photon Source,
Argonne National Laboratory, Argonne, Illinois 60439, USA}
\author{A. Paramekanti}
\affiliation{Department of Physics, University of Toronto, Toronto, Ontario M5S~1A7, Canada}
\author{Young-June Kim}
\email{yjkim@physics.utoronto.ca}
\affiliation{Department of Physics, University of Toronto, Toronto, Ontario M5S~1A7, Canada}
\date{\today}

\begin{abstract}
We report resonant inelastic X-ray scattering (RIXS) measurements on ordered double perovskite samples containing Re$^{5+}$ and Ir$^{5+}$ with $5d^2$ and $5d^4$ electronic configurations respectively.  In particular, the observed RIXS spectra of Ba$_2$YReO$_6$ and Sr$_2$MIrO$_6$ (M=Y, Gd) show sharp intra-t$_{2g}$ transitions, which can be quantitatively understood
using a minimal `atomic' Hamiltonian incorporating spin-orbit coupling ($\lambda$) and Hund's coupling ($J_H$). Our analysis yields
$\lambda=0.38(2)$eV with $J_H=0.26(2)$eV for Re$^{5+}$, and
$\lambda=0.42(2)$eV with $J_H=0.25(4)$eV for Ir$^{5+}$.  Our results provide the first sharp estimates for the Hund's coupling in $5d$ oxides, and
suggest that it should be treated on equal footing with spin-orbit interaction in multi-orbital $5d$ transition metal compounds.
\end{abstract}
\maketitle

\section{Introduction}

Hund's coupling, $J_H$, represents the local spin exchange interaction for electrons in multiorbital systems, and it is responsible for a variety of interesting phenomena in solids.
For example, Hund's coupling is responsible for spin-state transitions as a function of temperature in certain insulating $3d$ transition metal compounds \cite{Kahn44,CaCoO3,Yamaguchi1996,Radaelli2002,Vogt2000}. More remarkably, Hund's coupling has two distinct and contrary effects in multiband metals \cite{Luca2011}. On the one hand, it suppresses the atomic charge gap, making it energetically unfavorable for electrons to be localized and become a Mott insulator. On the other hand, it promotes strongly correlated bad-metal behavior by rendering Fermi liquid quasiparticles incoherent. This dichotomous role played by $J_H$ is now recognized to be important in the widely studied iron pnictides \cite{Hu_FeSc,Haule_FeSc,Johannes_FeSc,Yang_FeSc,Zhou_FeSc,Raghuvanshi_FeSc,Schafgans_FeSc}, as well as in ruthenates like
Ca$_2$RuO$_4$\cite{Georges2013,Ca2RuO4ARPES}.

In recent years, there has been an increasing interest in complex $5d$ oxides. In these systems, there is an
intricate interplay of electronic correlation, Hund's coupling, spin-orbit coupling (SOC) $\lambda$, and electron kinetic energy, which
leads to novel ground states \cite{SOCreview}. This underscores the need to accurately determine these energy scales, which has
important ramifications for magnetism, bad metal behavior, and Mott transitions in $5d$ oxides.
For example, density functional theory calculations predict the honeycomb material
(Na,Li)$_2$IrO$_3$ to exhibit a large bandwidth and weakly correlated behavior \cite{Mazin,Foyevtsova}. However, experiments show that it is better described as a
$J_{\rm eff}\!=\! 1/2$ Mott insulator \cite{Singh2010,Gretarsson2013,Chun2015}, in agreement with a recent exact diagonalization study which accounts for local correlation effects \cite{BHKim}.
Similarly, $d^4$ systems with strong SOC have been predicted to behave
as localized $J_{\rm eff}\!=\! 0$ insulators, with magnetism induced by exciton condensation \cite{Khaliullin2013,Akbari2014,Trivedi2015,Trivedi2017} or
impurity effects \cite{Dey},
while band theory \cite{Bhowal2015}  provides an itinerant magnetism explanation for the observed Ir magnetic moment in A$_2$YIrO$_6$ (A=Ba,Sr)
\cite{GCao_SYIO,Dey}.
Incorporation of correlation effects appear to be necessary to resolve the controversy in understanding magnetism in $d^4$ double perovskites.
The interplay of SOC and Hund's coupling is also clearly important in understanding the electronic ground states in multi-electron
$5d^2$ rhenates and $5d^3$ osmates \cite{GChen2011,Calder}.
While Hund's coupling is irrelevant for the single-hole atomic configuration of Mott insulating $5d^5$ iridates, it is important
for superexchange processes which involve  intermediate $5d^4$ configurations (two-hole states).
This determines the strength of the conventional Heisenberg interaction relative to the unconventional
Kitaev exchange which can drive an exotic quantum spin liquid in honeycomb-based materials  \cite{Jackeli2009,AIrO3,Singh2012,Gretarsson2013,Modic2014,Chun2015,bLi2IrO3,gLi2IrO32,Knolle2014,Sizyuk2014,Kimchi2015}.

Remarkably, despite this wide interest in complex $5d$ oxides and the importance of the Hund's coupling for understanding their magnetic properties, there has been
no direct and accurate experimental determination of $J_H$ in these systems.
The values for $J_H$ used in numerical calculations on $5d$ oxides vary widely, ranging from $J_H\!=\! 0.2$eV to $0.6$eV
\cite{Yamaji, BHKim, Foyevtsova,Dey, Pajskr}, while
analytical studies typically focus on the simple limits $J_H \gg \lambda$ or $J_H \ll \lambda$.

In this paper, we use resonant inelastic X-ray scattering (RIXS) to explore local spin-orbital excitations in Ir$^{5+}$ ($5d^4$) and Re$^{5+}$ ($5d^2$) double perovskites.
Use of the two complementary $5d$ insulating oxides modeled by an effective `atomic' Hamiltonian allows us to determine these important energy scales, $J_H$ and $\lambda$ with high precision.
We find $\lambda({\rm Ir})\!=\! 0.42(2)$eV with $J_H({\rm Ir})\! = \! 0.25(4)$eV, and
$\lambda({\rm Re}) \!=\! 0.38(2)$eV with $J_H ({\rm Re}) \!=\! 0.26(2)$eV. The $J_H$ values
obtained here represent the first measurements for rhenates and iridates.

\section{Experimental methods}

In our study, we choose to work with ordered
double perovskite (DP) compounds A$_2$BB'O$_6$ (B=Ir, Re) which offer two distinct advantages.
In Ba$_2$YReO$_6$ and Sr$_2$YIrO$_6$, the Re/Ir octahedra form a rock-salt structure with adjacent octahedra centered around inert Y$^{3+}$ ions.
The intervening electronically inactive YO$_6$ octahedra ensures that the overlap between the neighboring Re/Ir orbitals is small,
leading to extremely narrow spectral bandwidths as shown in our RIXS data.
This allows us to focus on the local physics, and justifies our use of an `atomic' Hamiltonian to model the data.
The second benefit of using DPs has to do with suppression of the
Jahn-Teller (J-T) instability. In a perovskite structure with octahedra formed by $d^2$ or $d^4$ ions, there is a tendency for a cooperative J-T effect, in which neighboring octahedra distort in a complementary manner which strongly breaks the local octahedral symmetry. However, if the J-T active octahedron is surrounded by octahedra containing non-J-T ions (such as Fe$^{3+}$ or Y$^{3+}$),
this instability is suppressed \cite{Oikawa}. As a result, although the ReO$_6$ and IrO$_6$ octahedra in the DP structure may undergo small rotations, they lead to very weak
deviations from an ideal local octahedral environment.

Two different experimental setups were used for the RIXS experiments at the Advanced Photon Source.
For the Ir $L_3$ (Re $L_2$) edge RIXS experiments carried out at the 9ID (27ID) beamline, the beam
was monochromatized by Si(844) [Si(400)] crystals. A spherical (1-m-radius) diced Si(844) [Si(773)]
analyzer was used to select final photon energy. In order to minimize the elastic background intensity, measurements
were carried out in a horizontal scattering geometry, for which the scattering angle $2 \theta$ was close to
90 degrees. The overall energy resolution of about 40 meV (FWHM) for Ir and 100 meV for Re
was obtained. The Re-DP samples used in our measurements,  Ba$_2$YReO$_6$, Ba$_2$FeReO$_6$, and Ca$_2$FeReO$_6$, were all polycrystalline powder samples pressed into pellets. For Ir-DP measurements, we used single crystals of Sr$_2$YIrO$_6$ and Sr$_2$GdIrO$_6$. The synthesis and characterization of these samples have been
previously reported \cite{GCao_SYIO,Kato_ReDP,oikawa_CFRO,AFRO_synthesis_ref1,AFRO_synthesis_ref2,Aharen_BYRO}, and all samples show high degree of B/B' order due to the valence difference.

\section{Experimental Results}

\begin{figure}[htbp!]
	\begin{center}
		\includegraphics[width=0.5\textwidth]{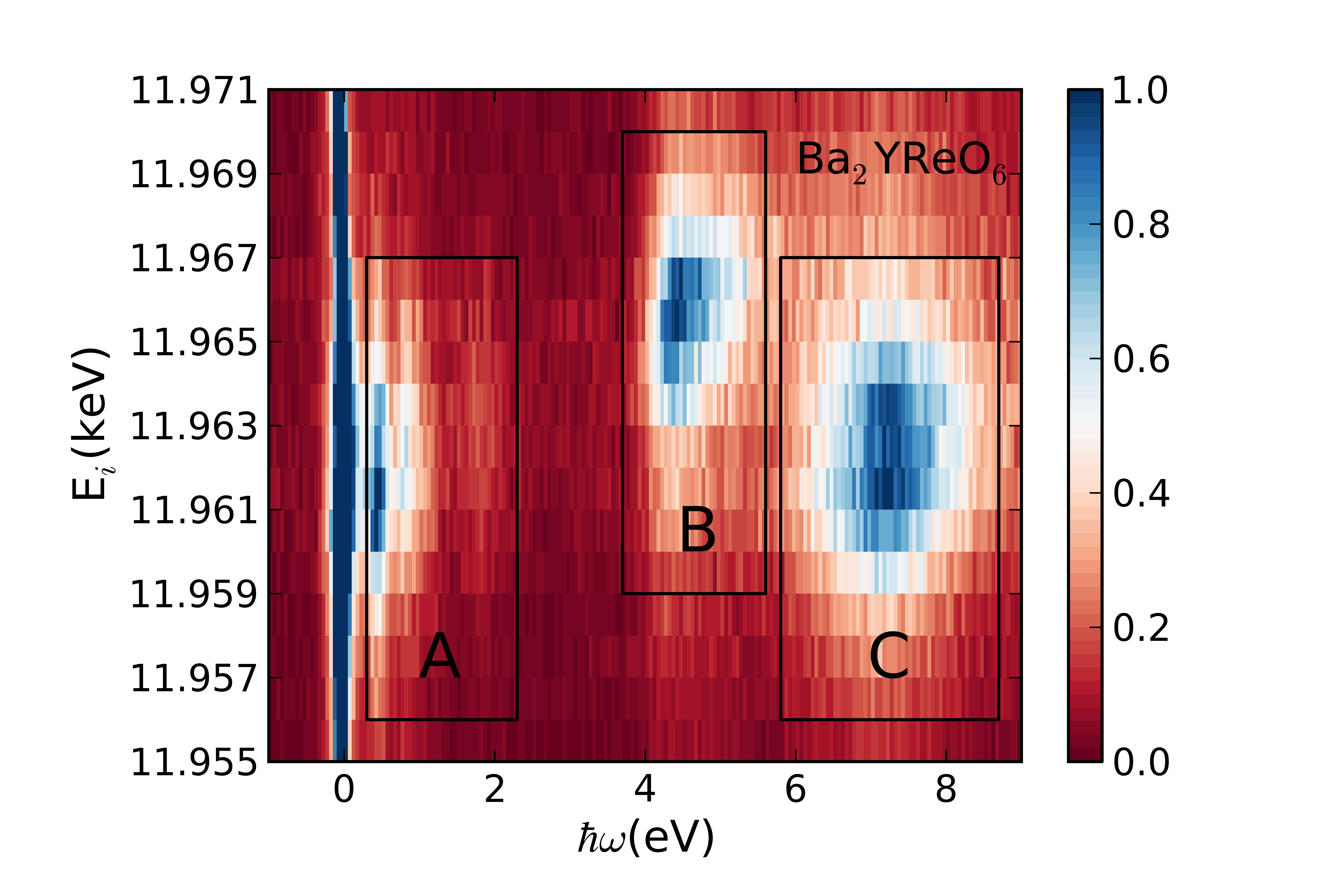}
		\caption{Incident energy (E$_i$) dependence of Ba$_2$YReO$_6$ RIXS spectra. The RIXS intensity is plotted as a function of incident energy E$_i$ (vertical axis) and energy transfer $\hbar\omega$ (horizontal axis). An arbitrary intensity scale is used where blue (red) denotes higher(lower) intensity.}
		\label{supp1}
	\end{center}
\end{figure}

Incident energy (E$_i$) dependence of Ba$_2$YReO$_6$ RIXS spectra are shown in Fig.~\ref{supp1}. One can resolve three main features with $\hbar\omega\lesssim$2.5eV (feature A), 4eV$\lesssim\hbar\omega\lesssim$6eV (feature B) and 6eV$\lesssim\hbar\omega\lesssim$8eV (feature C). Both features A and C show enhancement when $E_i$ is tuned near the resonance energy of E$_i\approx$11.961keV, whereas feature B resonates at slightly higher E$_i\approx$11.965keV. RIXS follows a second order process (dipole transition from 2p to 5d and another transition back to 2p)  with an intermediate state consisting of a $2p$ core hole and an excited electron in either t$_{2g}$ or e$_g$ states. Different resonant energies thus reflect different intermediate states in these transitions. This allows us to assign A and C as intra t$_{2g}$ and charge transfer (CT) excitation from the surrounding ligands to t$_{2g}$ states, respectively, and B as $t_{2g}-e_g$ transition. The intermediate states of both intra-t$_{2g}$ and CT excitation are $\underline{2p}t_{2g}^3$, where the underline denotes a $2p$ core-hole. On the other hand, the intermediate state for $t_{2g}-e_g$ transition is $\underline{2p}t_{2g}^2e_g^1$, which occurs at higher energy than $\underline{2p}t_{2g}^3$. The difference in resonant energies thus corresponds to the $t_{2g}-e_g$ splitting. As discussed earlier, the spatial extent of the $5d$ orbital leads to a large $t_{2g}$-$e_g$ splitting, while the $t_{2g}$ orbitals are further split by $J_H$ and $\lambda$ as shown in the figure. We note that strong fluorecence features were observed around 10~eV in the study of metallic rhenate samples ReO$_2$ and ReO$_3$ \citep{Smolentsev}, which is absent in our $E_i$-dependence study of insulating Ba$_2$YReO$_6$. Qualitatively similar incident energy dependence has been reported for iridate samples in the past \cite{Ishii2011}

\begin{figure*}[tb]
	\begin{center}
		\includegraphics[width=0.45\textwidth]{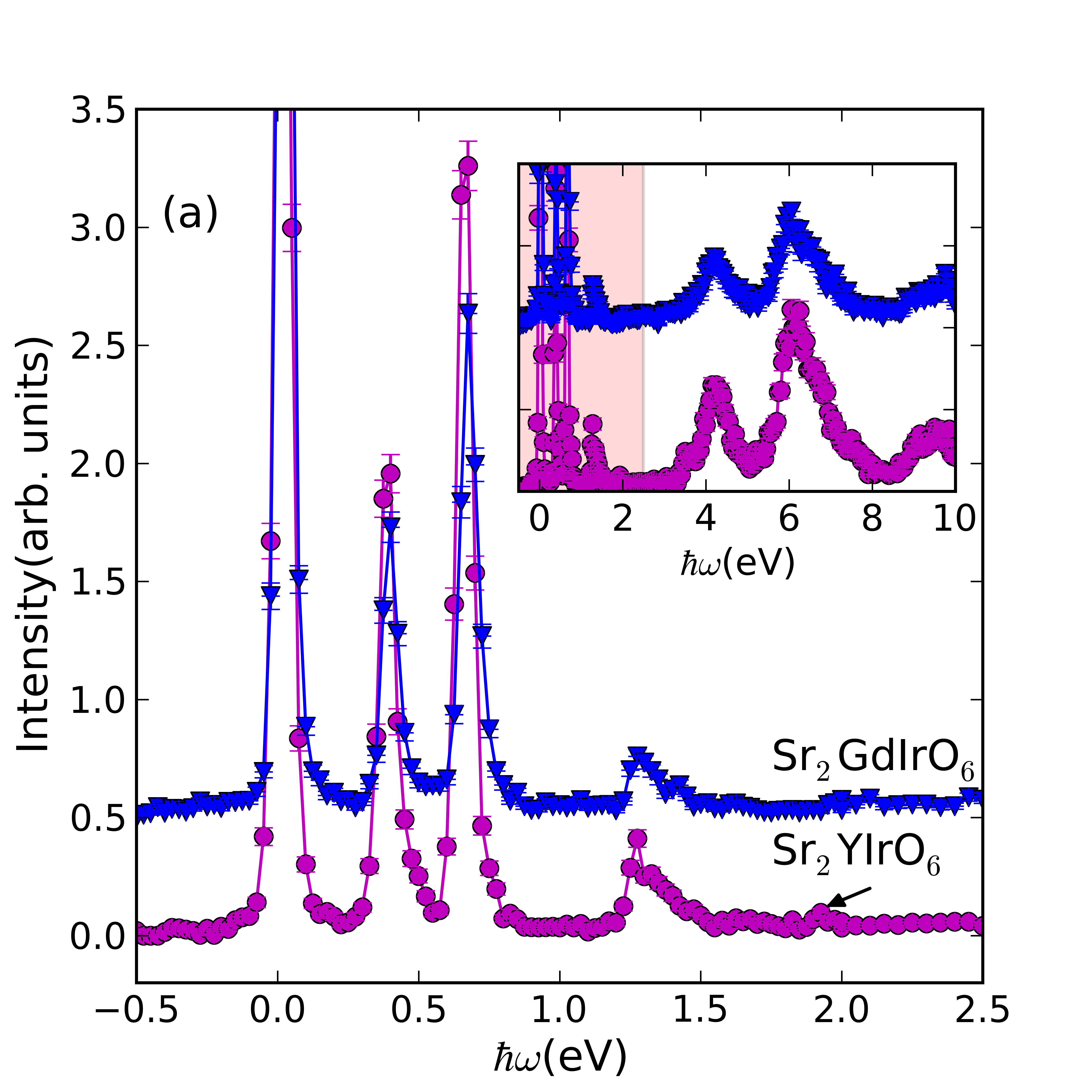}%
		~
		\includegraphics[width=0.45\textwidth]{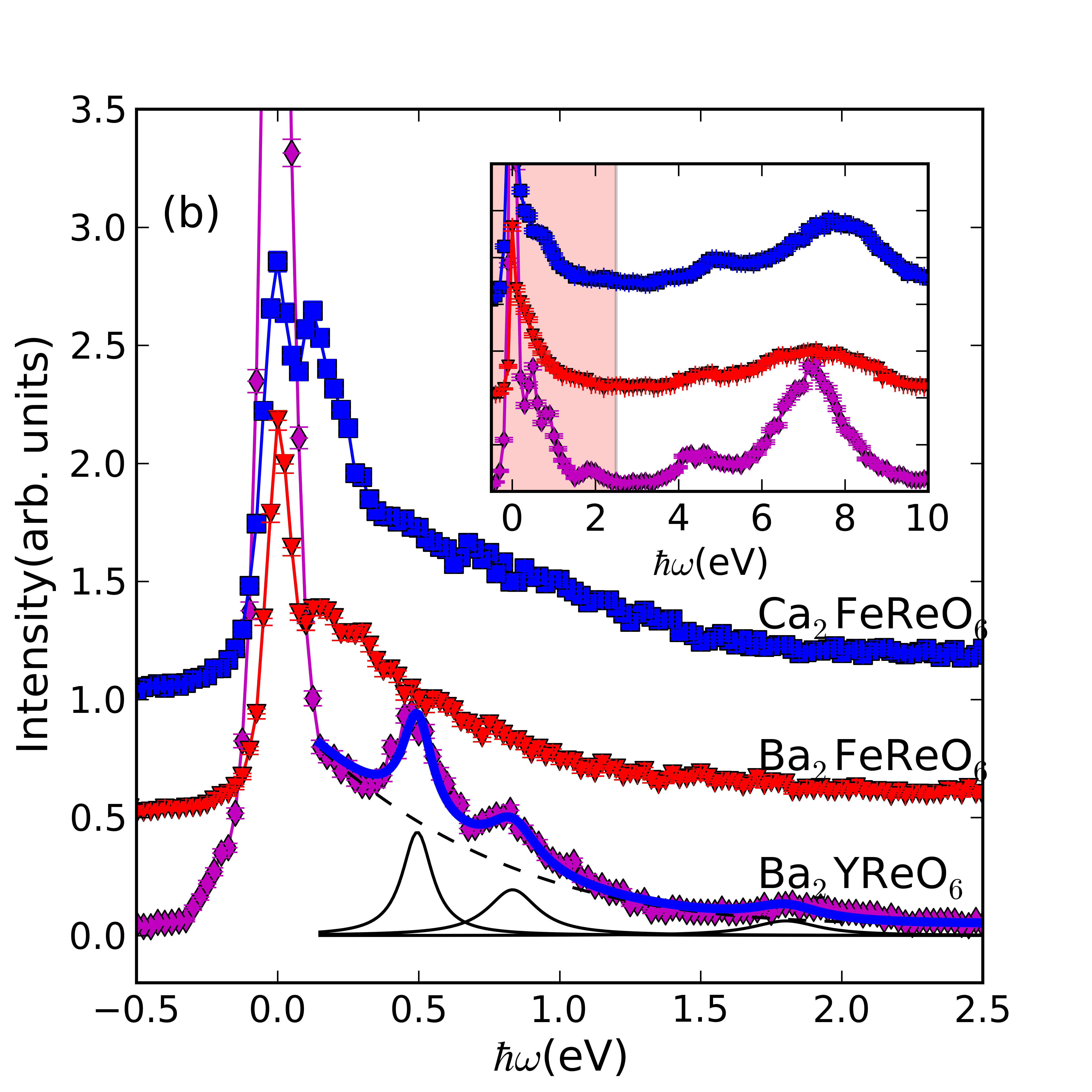}
		\caption{RIXS spectra of (a) Ir double perovskites and (b) Re double perovskites. Main panels show details of intra-t$_{2g}$ excitations in the energy range $\hbar\omega<2.5eV$, while full RIXS spectra covering wide range of energy transfer $\hbar\omega<$10eV are shown in insets. Incident energy E$_i$=11.215keV and E$_i$=11.961keV with fixed Q near $2 \theta=90^\circ$ were used to obtain spectra in (a) and (b) respectively. The scans are vertically offset for visual clarity, and the intensity scale is arbitrary. The arrow in (a) indicates the weak $\sim$2eV feature (see text). The thick blue line in (b) is a fit to the Ba$_2$YReO$_6$ spectrum as described in the text. Contributions from individual peaks are shown as black solid lines. Dashed line indicates the sloping quartic background.}
		\label{RIXS spectra}
	\end{center}
\end{figure*}

Room temperature RIXS spectra of all DP samples are shown in Fig.~\ref{RIXS spectra}. Wide-range scans are shown in insets of panels (a) and (b), for Ir-DP and Re-DP, respectively. The low-energy region
below $2.5$eV are zoomed in and shown in the main panels. All samples are found to exhibit qualitatively
similar excitation spectra: a set of sharp peaks in the low energy range $\lesssim 2.5$eV (the sharp $\hbar \omega =0$ peaks are due to elastic background and represent instrumental resolution),
and two broader peaks in the high energy $4$-$8$eV range. Despite the similarity in the peak positions, we find a systematic difference in the peak widths when comparing different samples.
In particular, metallic Ba$_2$FeReO$_6$ and semimetallic Ca$_2$FeReO$_6$ exhibit very broad features, clearly contrasting the other (insulating) samples, which exhibit well-resolved peaks.
We thus focus below on the spectra of only the insulating samples in order to extract quantitative information from the peak positions.

As shown in Fig.~\ref{RIXS spectra} (a), the Ir-DP samples display sharp features that are resolution limited. The three inelastic peak positions in both Sr$_2$GdIrO$_6$ and Sr$_2$YIrO$_6$ can be read directly from their spectra: $0.39(2)$eV, $0.66(2)$eV, and $1.30(6)$eV. No momentum dependence was found for these features (See Appendix). In addition, there is a very weak feature at $\sim 2$eV in both iridates. We note that the crystal structure of these two compounds are different; Sr$_2$YIrO$_6$ and Sr$_2$GdIrO$_6$ crystallize, respectively, in monoclinic and cubic symmetry \cite{GCao_SYIO}. In addition, the IrO$_6$ octahedra are slightly flattened along the apical direction in Sr$_2$YIrO$_6$, with distinct Ir-O bond lengths $1.9366\AA$ (apical), $1.9798\AA$, and $1.9723\AA$. However, the octahedra in
Sr$_2$GdIrO$_6$ are almost undistorted \cite{GCao_SYIO}. The lack of momentum dependence of the inelastic features and the fact that we observe almost the same peak positions in these two systems suggest that the electronic structure is determined by local physics such as $\lambda$ and $J_H$, and is unaffected by the global symmetry or the presence of a small distortion.

In contrast to the Ir-DPs, the spectral features in the Re-DPs are much broader, partly because of coarser energy resolution. In addition, metallic samples are expected to exhibit large peak width resulting from stronger damping due to particle-hole continuum as well as powder averaging effect, as is seen for Ba$_2$FeReO$_6$ and Ca$_2$FeReO$_6$. However, for insulating Ba$_2$YReO$_6$, we find three peaks that can be clearly resolved on top of a broad continuum, so we focus on only this rhenate in our
analysis below. The low energy continuum is modeled with a quartic background - as discussed later, we tentatively attribute this background to coupled multi-phonon/magnon contributions. To extract peak positions, the low energy spectrum from $0.15$-$2.5$eV is fitted with 3 Lorentzians as shown in Fig.~\ref{RIXS spectra}(b). From these fits, we extract peak positions $0.49(3)$eV, $0.83(4)$eV and $1.85(5)$eV. The corresponding FWHM are $0.13(4)$eV, $0.22(8)$eV, and $0.29(8)$eV, respectively.

\section{Theoretical model}

We next turn to a theoretical modelling of our data. We begin by noting that the sharp, momentum-independent, inelastic peaks found in our RIXS measurements suggest that a local Hamiltonian is appropriate for understanding these excitations. Furthermore, the two sets of compounds in our RIXS study are particle-hole conjugates,
with the rhenates being at a filling of $2$-electrons while the iridates are at a filling of $2$-holes. While the local atomic interactions are particle-hole symmetric, SOC breaks this symmetry. As a result, projecting to the $t_{2g}$ orbitals, both sets of materials can be described
by the same Kanamori Hamiltonian,
\begin{equation}
H_{\rm eff}=-2 J_H \vec{S}^2 - \frac{J_H}{2} \vec{L}^2 \pm \lambda(\vec{l}_1\cdot\vec{s}_1+\vec{l}_2\cdot\vec{s}_2)
\end{equation}
where $+(-)$ with the $2$-hole ($2$-electron) picture
applies for the $d^4 (d^2)$ configuration, and $\vec L$ and $\vec S$ refer to the total orbital and spin angular momenta respectively of the two particles.
For $J_H \ll \lambda$, the eigenstates are obtained by perturbing around the noninteracting limit which corresponds to occupying the $j_{\rm eff}=1/2$ and $j_{\rm eff}=3/2$ multiplets arising from SOC \cite{endnote44}. For large $J_H$, the eigenstates should be understood as arising from $\vec S$ and $\vec L$ being locked together by SOC.
In either limit, the $d^2$ ($d^4$) case exhibits a ground state with $J_{\rm eff}=2$ ($J_{\rm eff}=0$). The $d^2$ vs $d^4$ difference arises due to the
opposite signs of the effective SOC.
While we expect $J_H$ and $\lambda$ to be similar for Ir and Re, we do not demand that they be identical.

\begin{figure*}[tb]
	\begin{center}
		\includegraphics[width=\textwidth]{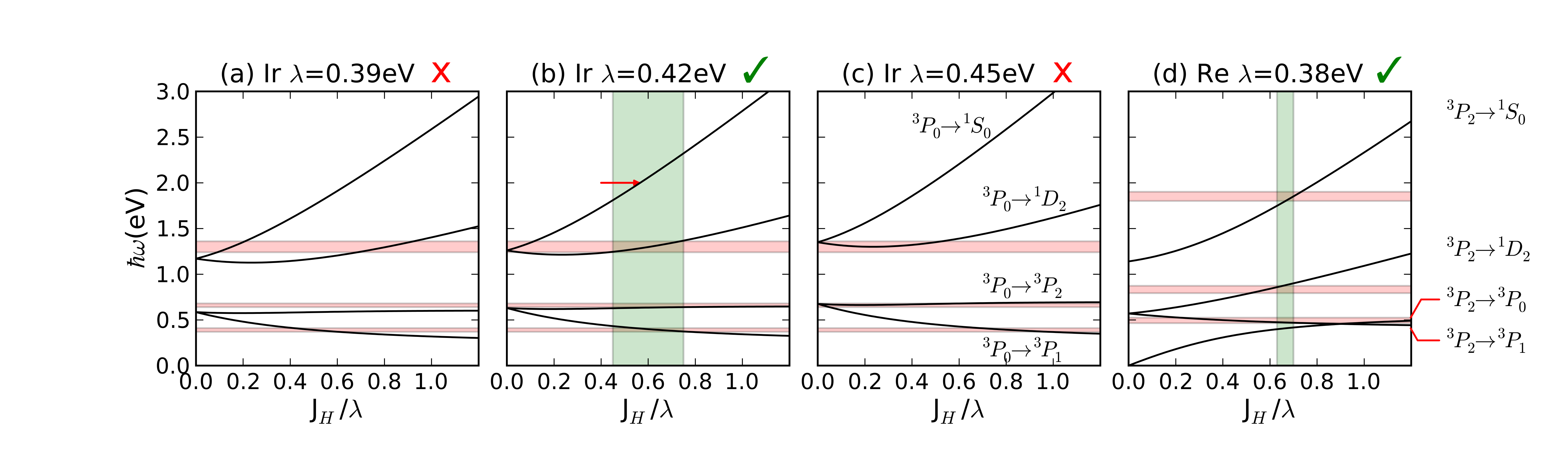}
		\caption{Calculated excitation energies as a function of $J_H/\lambda$ are plotted as solid black lines, for Ir (a-c) and Re (d) double perovskites for indicated
		$\lambda$ values. Experimentally determined excitation energies (Ir-DP: $0.39(2)$eV, $0.66(2)$eV, and $1.30(6)$eV; Re-DP: $0.49(3)$eV, $0.83(4)$eV, $1.85(5)$eV) are plotted as horizontal colored bands (pink), whose widths reflect the experimental uncertainty. States involved in these transitions are labelled using nomenclature in the $J_H\gg\lambda$ limit as $^{2S+1}L_J$. (a,c) and (b,d) show incorrect and correct choices of $\lambda$ as described in the text, which are indicated by $\bf{\times}$ and $\bf{\surd}$ respectively. Green shaded regions in (b) and (d) illustrate $J_H/\lambda$ values for which {\it all} the observed modes intersect the calculated curves. Red arrow in (b) denotes position of the weak feature $\sim 2$eV in the Ir-DPs which would correspond to the highest computed excitation energy.}			
		\label{theory}
	\end{center}
\end{figure*}

To extract $\lambda$ and $J_H$, we plot the calculated excitation energies for the two cases (Ir and Re) as a function of
$J_H/\lambda$ for different choices of
$\lambda$, as shown in Fig.~\ref{theory}, and superpose on this the observed peak positions. For the correct choice of $\lambda$, the computed curves should intersect {\it all}
the observed peaks at a {\it common} value of $J_H/\lambda$, allowing us to extract both $\lambda$ {\it and} $J_H/\lambda$.

Figs.~\ref{theory}(a-c) show the theoretically computed spectra for
the Ir-DPs as a function of $J_H/\lambda$ for increasing values of SOC: $\lambda=0.39$eV, $0.42$eV, and $0.45$eV respectively. We also show on these plot the three experimentally observed modes as
thick colored lines (pink), with the width indicating the experimental uncertainty. For $\lambda=0.39$eV in Fig.~\ref{theory}(a), we find that the central mode does not intersect the computed spectra for any choice of $J_H/\lambda$, while the highest and lowest energy modes intersect for $0.65\lesssim J_H/\lambda\lesssim0.9$ and
$0.45 \lesssim J_H/\lambda\lesssim 0.7$ respectively. For $\lambda=0.42$eV in Fig.~\ref{theory}(b), we show that there is a range $0.5 \lesssim J_H/\lambda \lesssim 0.7$, demarcated
by the green shaded region, over which {\it all} observed modes intersect the theoretical curves.
Note that Fig.~\ref{theory}(b) also marks the location of the weak $\sim 2$eV mode with an arrow showing that this also occurs
in the correct regime of $J_H/\lambda$; however, given the low intensity of this mode, it
should only be viewed as a consistency check. Finally, for even larger SOC, $\lambda=0.45$eV in Fig.~\ref{theory}(c), we
find that while the central mode intersects the computed spectra over a
wide range of $J_H/\lambda$, the highest and lowest modes now intersect the theoretical curves for nonoverlapping regimes $J_H/\lambda \lesssim 0.5$ and
$0.7\lesssim J_H/\lambda \lesssim 1.05$ respectively.
Thus, there is no single choice of $J_H/\lambda$ which would explain all the observed modes for the cases in Fig.~\ref{theory}(a) and (c), while $\lambda=0.42$eV in Fig.~\ref{theory}(b) is a viable choice for the SOC. We show a similar plot for the Re-DP in Fig.~\ref{theory}(d) for a choice $\lambda=0.38$eV, where we find a small common
intersection window near $J_H/\lambda \approx 0.7$.
Using this procedure, we conclude that the range of $\lambda$ values over which such common intersections occur provides an estimate of the SOC, while the
window of the common intersection region yields an estimate of $J_H/\lambda$.
A least squares fit to the peak positions
allows us to determine $J_H$ and $\lambda$ with remarkably high precision: $\lambda({\rm Ir})\!=\! 0.42(2)$eV with $J_H({\rm Ir})\! = \! 0.25(4)$eV, and
$\lambda({\rm Re}) \!=\! 0.38(2)$eV with $J_H ({\rm Re}) \!=\! 0.26(2)$eV.
Our result for $\lambda({\rm Ir})$ is consistent
with previous experiments on the single-hole $5d^5$ iridates  \cite{Liu2012,Hozoi2014,Sala2014,Sala2014PRB}.
Further, since Re ($Z \!=\! 75$) is close to Ir ($Z\!=\! 77$) in the periodic table, we expect similar values for $\lambda$ and $J_H$, with a smaller $\lambda$ for Re given its lower $Z$,
as is borne out by our analysis. Our work highlights the need to treat $J_H$ and $\lambda$ on equal footing in complex $5d$ oxides.

Interestingly, our model also leads to a simple explanation for why the higher energy peaks in the RIXS data in the $1$-$2$eV range
(Fig.~\ref{RIXS spectra}) have much smaller spectral weight than the two lower energy inelastic
peaks. As seen from the theoretical plots in Fig.~\ref{theory}, at $J_H/\lambda\!=\! 0$, the iridates (rhenates) have two
sets of excitations, which correspond to exciting one or two holes (electrons) from
$j_{\rm eff}\!=\! 1/2 \to 3/2$ ($j_{\rm eff}\!=\! 3/2 \to 1/2$). These occur at
excitation energies $3\lambda/2$ and $3\lambda$ respectively. However, the latter two-particle excitation is not accessed
within the RIXS process at $J_H/\lambda\!=\! 0$, and thus has {\it zero} spectral weight. Turning on a small $J_H\!>\!0$ modifies this result in two important ways:
(i) it splits these excitations into multiple branches as seen from Fig.~\ref{theory}, and (ii) it leads to a small nonzero spectral weight $\sim (J_H/\lambda)^2$
for the higher energy peaks from interaction-induced mixing between the  $j_{\rm eff}\!=\! 1/2$ and $j_{\rm eff}\!=\! 3/2$ levels. We have confirmed this picture
with a theoretical calculation of the RIXS spectrum for the iridate samples.

\section{Theoretical calculation of RIXS spectrum}

The Kramers-Heisenberg expression for the two-photon scattering cross section is given by
\bea
\frac{d^2\sigma}{d\Omega d\omega} &=& \frac{\omega'}{\omega} \sum_f
\left| \sum_n \frac{\la f| T^\dagger | n \ra \la n | T | i \ra}{E_i - E_n + \hbar \omega + i \frac{\Gamma_n}{2}} \right|^2 \nonumber \\
&\times& \delta(E_i - E_f + \hbar \omega - \hbar \omega').
\eea
Here, $i,n,f$ refer to initial, intermediate, and final states respectively with energies $E_i, E_n, E_f$, and $\Gamma_n$ is the inverse lifetime
of the intermediate state with a core-hole.
$\omega, \omega'$ are the incoming and outgoing photon frequencies. The transition is induced by the dipole operator $T \sim \hat{\epsilon} \cdot \br$,
where $\hat{\epsilon}$ denotes the photon polarization. Here, we focus on the $d^4$ iridates at the $L_3$ resonance within the hole picture,
for which the initial and final states
come from the two-hole eigenstates on Ir with spin-orbit coupling and Hund's interaction, while the intermediate state corresponds to a single core hole
in the atomic $2P_{3/2}$ manifold and a single hole in the $j_{\rm eff}=1/2$ manifold. On resonance, with $\omega' \approx \omega$ (since the energy transfer
is much smaller than the incoming or outgoing photon energies), the cross section simplifies to
\begin{widetext}
\bea
\frac{d^2\sigma}{d\Omega d\omega} &\approx&  \left|\frac{1}{E_i - \bar{E}_n + \hbar \omega + i \frac{\bar{\Gamma}_n}{2}} \right|^2 \sum_f
\left| \sum_n \la f| T^\dagger | n \ra \la n | T | i \ra \right|^2
\delta(E_i \!-\! E_f \!+\! \hbar \omega - \hbar \omega')
\eea
\end{widetext}
where $\bar{E}_n, \bar{\Gamma}_n$ are the average energy and inverse lifetime of the intermediate states.

We can further simplify the transition matrix element as
\bea
\la n | T | i \ra &=& \epsilon^\alpha_{\rm in} \la n | p^\dg_{\beta\sigma}  d^\pdg_{\alpha\beta\sigma}  | i \ra\\
\la f | T^\dagger | n \ra &=& \epsilon^\mu_{\rm out} \la f | d^\dg_{\mu\nu\sigma'} p^\pdg_{\nu\sigma'} | n \ra
\eea
where we have restricted attention to parity-allowed nonzero dipole matrix elements. Here $p^\dg_{\alpha\sigma}$ creates a $2P$ core-hole in orbital
$\alpha$ (i.e., $p_x,p_y,p_z$) with spin $\sigma$, while $d^\dg_{\alpha\beta\sigma}$ creates a $d$-hole in the $t_{2g}$ orbital (i.e., $d_{yz},d_{zx},d_{xy}$)
with spin $\sigma$. Based on the experimental
set-up, we fix the incoming polarization to be along the cubic $x$-axis, and average the outgoing polarization within the $yz$ plane since the
scattering geometry fixes $\hat{\epsilon}_{\rm in} \cdot \hat{\epsilon}_{\rm out}=0$. Using exact diagonalization
for the Hilbert space consisting of $15$ states for the two-hole problem with the Hamiltonian in Eq.~(1),
and the single-hole eigenstates of the $2P_{3/2}$ and $j_{\rm eff}=1/2$, we
obtain the theoretical RIXS spectrum. Fig.~\ref{supp3} below
shows an example of the theoretical spectrum obtained by convolving the above theoretical expression with a Lorentzian resolution function with an experimentally
determined width $\sim 40$meV, for a choice $\lambda=0.42$eV and $J_H=0.25$eV. We find that the
two lower energy peaks have a strong intensity since they emerge from the allowed single-particle transition across the spin-orbit gap $3\lambda/2$, while the
two higher energy peaks have a much lower intensity which scales as $\sim (J_H/\lambda)^2$ for small interactions since they emerge from exciting {\it two} holes across the
spin-orbit gap which is forbidden in the absence of hole-hole interactions arising from Hund's coupling.
The resulting spectral intensities are in good agreement with our experimental results.
\begin{figure}[htbp!]
	\begin{center}
		\includegraphics[width=0.5\textwidth]{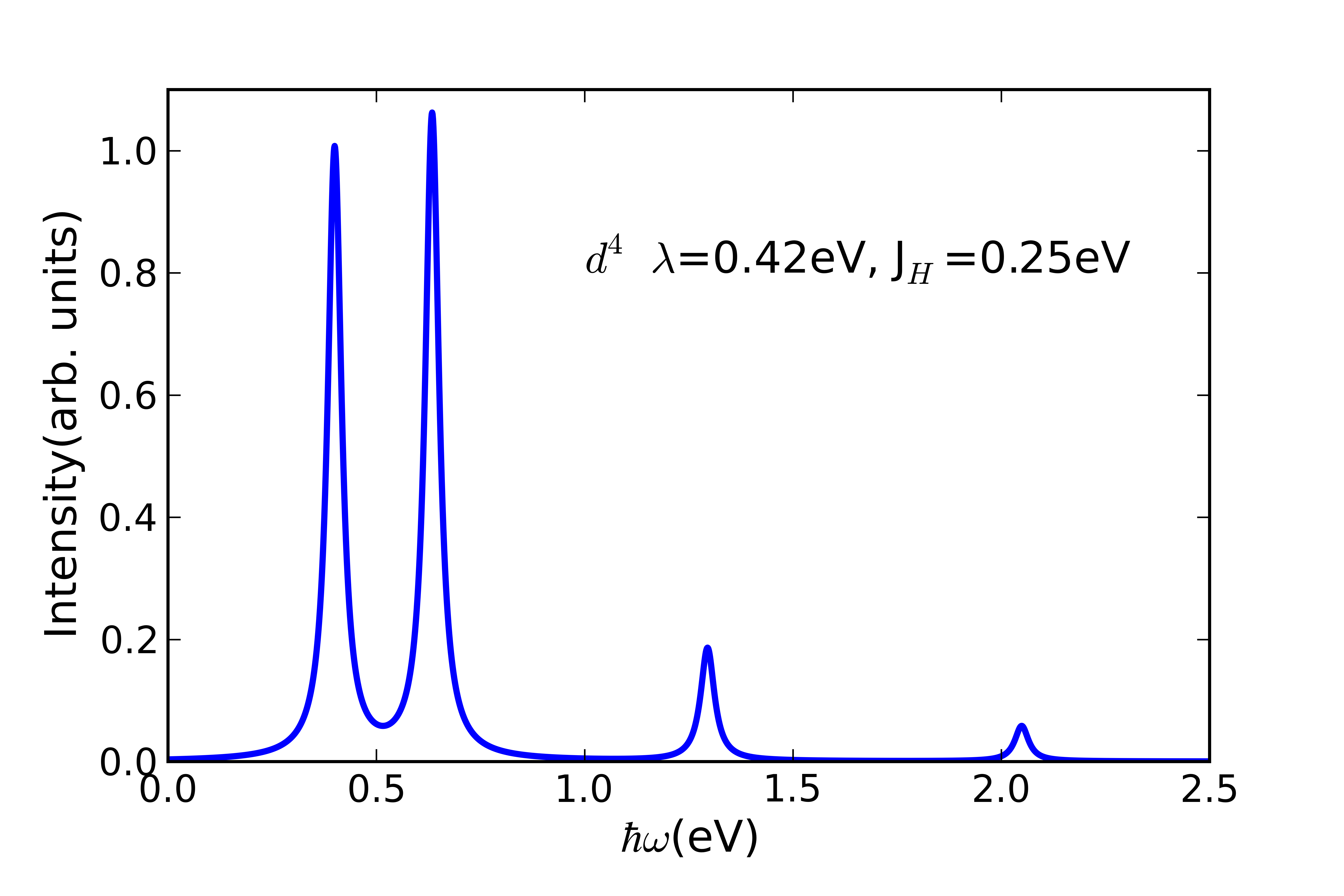}
		\caption{RIXS spectrum computed for $d^4$ iridates with $\lambda=0.42$eV, $J_H=0.25$eV.}
		\label{supp3}
	\end{center}
\end{figure}

\section{Discussion}

Despite the excellent agreement between theory and experiments in Fig.~\ref{theory}, there are two unresolved issues.
(i) For Re$^{5+}$, the lowest energy peak is expected to be at $\sim\! 0.4$eV. Although this peak is not observed as distinct from the $0.49$eV
peak in our data, it is possible that there are two nearby peaks which are not resolved in our experiment. (ii) For the Re-DPs, there is considerable
spectral weight in the low energy continuum below $\sim\! 0.3$~eV. The significant inelastic scattering intensity that was treated as sloping background in our fitting for
Ba$_2$YReO$_6$ (Fig.~\ref{RIXS spectra}(b)) remains even after subtracting off the non-resonant background (see Appendix).
While we can rule out magnon or phonon excitations for energies $\gtrsim 100$meV based on neutron scattering results \cite{Plumb_BFRO}, multiphonon excitations or some collective excitations of coupled degrees of freedom could exist in this energy range.
Future measurements with much higher energy resolution could address these issues.

In conclusion, our RIXS experiments on local spin-orbital excitations in Re and Ir double perovskites, together with a well-justified local model Hamiltonian, allows us to
reliably extract the SOC $\lambda \! \sim \! 0.4$eV and Hund's coupling $J_H \! \sim \! 0.25$eV for rhenates and iridates. We note that a recent study of the
5d$^3$ osmate Ba$_2$YOsO$_6$ reported a  smaller $\lambda\!=\! 0.32(6)$eV, and a larger $J_H\!= \!0.3(2)$eV \cite{Taylor}. Although large error bars make these values consistent with our results, it will be interesting to examine whether the discrepancy represents a real difference between $d^3$ and $d^2$/$d^4$ systems. Finally, our results are qualitatively consistent with a $J_{\rm eff}\!=\! 0$ ($J_{\rm eff}\!=\! 2$) for the ground state of the $d^4$ iridates ($d^2$ rhenates); however,
our finding that $J_H \!<\! \lambda$ might require revisiting theories of exotic magnetism in $d^2$ systems based on a strong coupling $J_H/\lambda \! \gg \! 1$ approach \cite{GChen2011}.

\acknowledgments

Research at the University of Toronto was supported by the Natural
Sciences and Engineering Research Council of Canada through
Discovery Grant. B.Y. would like to acknowledge support from the Canada Graduate Scholarships- Master's Program. B.C.J. and T.W.N. acknowledges the support by IBS-R009-D1. G.C. acknowledges NSF support via grants DMR 1265162 and DMR 1712101. This research used resources of the Advanced Photon Source, a U.S. Department of Energy (DOE) Office of Science User Facility operated for the DOE Office of Science by Argonne National Laboratory under Contract No. DE-AC02-06CH11357.

\appendix

\section{Background subtraction}

\begin{figure}[htbp!]
	\begin{center}
		\includegraphics[width=0.45\textwidth]{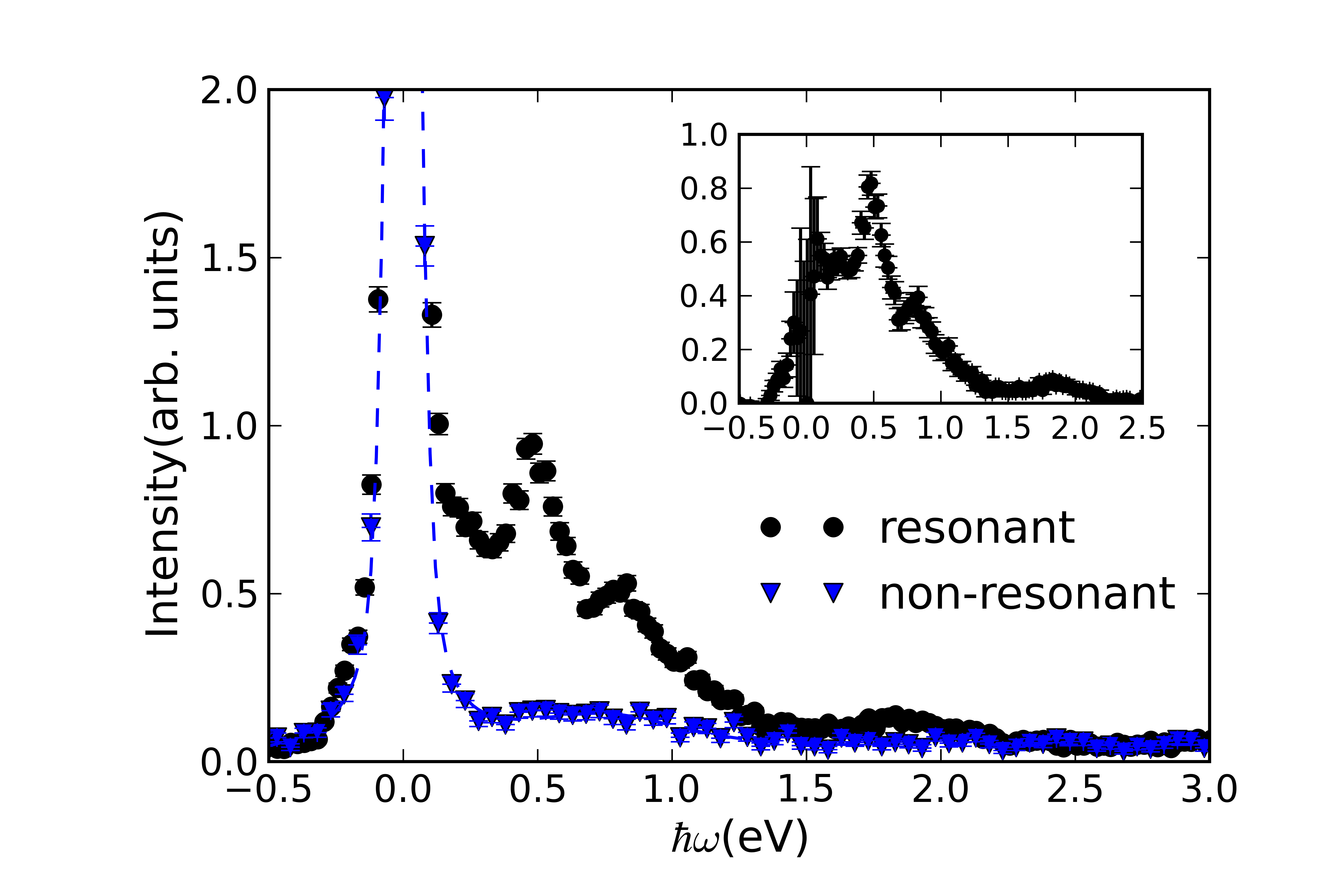}
		\caption{RIXS spectrum of Ba$_2$YReO$_6$ at resonant (E$_i$=11.961keV) and non-resonant incident energy (summed from E$_i$=11.969keV to E$_i$=11.971keV). The spectra have been scaled to have the same intensity at zero energy transfer. The blue dashed line is fit to the spectrum at non-resonant energy and is used as the background. \textit{Inset}: Inelastic features obtained by subtracting the background off the resonant spectrum. Clear spectral weight is observed for energy transfer $\hbar\omega<$0.3eV. }
		\label{supp2}
	\end{center}
\end{figure}

Since intensities of the low energy resonant inelastic features are greatly reduced for E$_i\gtrsim$11.969keV, we can use the spectra in this $E_i$ range as non-resonant background and subtract from our raw data to study the low energy excitations in Ba$_2$YReO$_6$ \citep{Ellis}. The non-resonant energy spectrum is obtained by summing over the spectra with E$_i$=11.969keV-11.971keV, and subtracted off from the spectrum at resonant energy (E$_i$=11.961keV) as shown in Fig.~\ref{supp2}.\citep{Ellis} The background subtracted spectrum shown in Fig.~\ref{supp2} inset clearly reveals the presence of significant spectral weight for $\hbar\omega\lesssim$0.3eV. This continuum, which exist in $\it{both}$ metallic Ba$_2$FeReO$_6$ and insulating Ba$_2$YReO$_6$, is not captured in our atomic model, and will require consideration of multiphonon or other collective excitations.

\section{S$\lowercase{r}_2$YI$\lowercase{r}$O$_6$: Q-dependence}
\begin{figure}[htbp!]
	\begin{center}
		\includegraphics[width=0.4\textwidth]{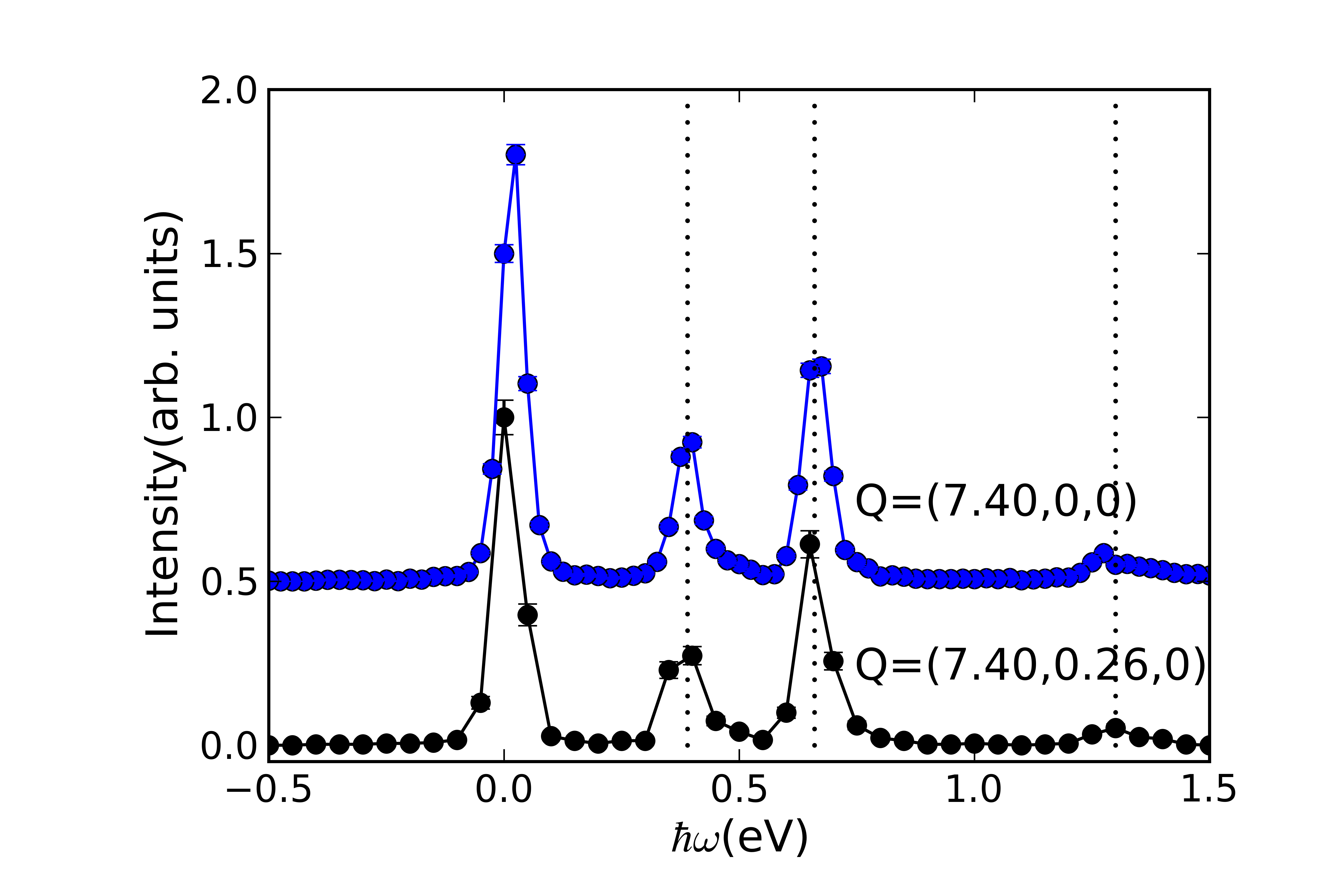}
		\caption{RIXS spectra of Sr$_2$YIrO$_6$ at $\bf{Q}$=(7.40,0,0) and $\bf{Q}$=(7.40,0.26,0) respectively. E$_i$=11.215keV was used in obtaining both spectra. An arbitrary intensity scale is used and the spectra have been shifted for visual clarity.}
		\label{supp3}
	\end{center}
\end{figure}

We show Sr$_2$YIrO$_6$ RIXS spectra measured at two different Q vectors, separated by approximately a quarter of the Brillouin Zone, in Fig.~\ref{supp3}. The inelastic features remain sharp and show no Q dependence, indicating the local nature of these excitations.


\begin{thebibliography}{58}
\bibitem[{end()}]{endnote44}
\bibinfo{note}{We use lower case $j_{\rm eff}$ to denote single particle(hole)
  states, while upper case $J_{\rm eff}$ is used for total angular momentum of
  an ion.}
\end{thebibliography}

\end{document}